\documentclass[11pt,twoside,a4paper,cmspaper,final,collab]{cms-tdr}
\def\svnVersion{54fe98c-D}\def\svnDate{2026/07/07}\def\cmsCernNoTag{CERN-EP-2026-124}\def\cmsCernDate{\today}\def\cmsMessage{Submitted to Physical Review Letters}
\begin{document}\cmsNoteHeader{EXO-23-001}

\newcommand{\pS}{\ensuremath{\text{S}}\xspace}
\newcommand{\mS}{\ensuremath{m_{\pS}}\xspace}
\newcommand{\pphi}{\ensuremath{\upphi}\xspace}
\newcommand{\mPhi}{\ensuremath{m_{\pphi}}\xspace}
\newcommand{\PApr}{\HepParticle{A}{}{\prime}\xspace}
\newcommand{\mApr}{\ensuremath{m_{\PApr}}\xspace}
\newcommand{\dark}{\ensuremath{\text{D}}\xspace}
\newcommand{\lambdaD}{\ensuremath{\Lambda_{\dark}}\xspace}
\newcommand{\alphaD}{\ensuremath{\alpha_{\dark}}\xspace}
\newcommand{\Ncdark}{\ensuremath{N_{c_{\dark}}}\xspace}
\newcommand{\TD}{\ensuremath{T_{\dark}}\xspace}
\newcommand{\pp}{\ensuremath{\Pp\Pp}\xspace}
\newcommand{\pipi}{{\ensuremath{\PGpp\PGpm}}\xspace}
\newcommand{\nSUEPConstituents}{\ensuremath{n_{\text{constituent}}^{\text{SUEP}}}\xspace}
\newcommand{\boostedSphericity}{\ensuremath{S^{\text{SUEP}}_{\text{boosted}}}\xspace}
\newcommand{\HTHLT}{\ensuremath{\HT^\text{HLT}}\xspace}
\newcommand{\scoutinglumi}{\ensuremath{127\fbinv}\xspace}
\newcommand{\corrF}{\ensuremath{c^{\text{bin}}}\xspace}
\newcommand{\BF}{\ensuremath{\mathcal{B}}\xspace}
\newcommand{\NSR}{\ensuremath{N_\text{SR}}\xspace}
\newcommand{\NSRbin}{\ensuremath{\NSR^\text{bin}}\xspace}
\newcommand{\NAa}{\ensuremath{N_\text{A}}\xspace}
\newcommand{\NB}{\ensuremath{N_\text{B}}\xspace}
\newcommand{\NC}{\ensuremath{N_\text{C}}\xspace}
\newcommand{\ND}{\ensuremath{N_\text{D}}\xspace}
\newcommand{\NE}{\ensuremath{N_\text{E}}\xspace}
\newcommand{\NF}{\ensuremath{N_\text{F}}\xspace}
\newcommand{\NGg}{\ensuremath{N_\text{G}}\xspace}
\newcommand{\NH}{\ensuremath{N_\text{H}}\xspace}
\newcommand{\NFbin}{\ensuremath{\NF^\text{bin}}\xspace}
\newcommand{\momi}{\ensuremath{p_{i}}\xspace}
\newcommand{\SAB}{\ensuremath{\mathcal{S}^{\alpha\beta}}\xspace}

\newlength\figWidth
\newlength\figWidthThree

\ifthenelse{\boolean{cms@external}}{%
	\setlength{\figWidth}{0.99\linewidth}
	\setlength{\figWidthThree}{0.32\textwidth}
	\newcommand{\appname}{End Matter}
	\newcommand{\EndMatterA}{the \appname}
	\newcommand{\EndMatterB}{the \appname}
	\newcommand{\cmsUpperLeft}{Left\xspace}
	\newcommand{\cmsUpperRight}{Middle\xspace}
	\newcommand{\cmsLower}{Right\xspace}
}{
	\setlength{\figWidth}{0.49\textwidth}
	\setlength{\figWidthThree}{0.49\textwidth}
	\newcommand{\EndMatterA}{Appendix~\ref{app:correction}\xspace}
	\newcommand{\EndMatterB}{Appendix~\ref{app:validation}\xspace}
	\newcommand{\cmsUpperLeft}{Upper left\xspace}
	\newcommand{\cmsUpperRight}{Upper right\xspace}
	\newcommand{\cmsLower}{Lower\xspace}
}

\cmsNoteHeader{EXO-23-001}
\title{Search for soft unclustered energy patterns in proton-proton collisions at \texorpdfstring{$\sqrt{s} = 13\TeV$}{sqrt(s) = 13 TeV} using data scouting}

\date{\today}

\abstract{
A search for soft unclustered energy patterns (SUEPs) is conducted using proton-proton collision data corresponding to an integrated luminosity of \scoutinglumi at a center-of-mass energy of 13\TeV, collected via the data scouting stream of the CMS experiment at the LHC. Only the results of the high-level trigger reconstruction are recorded to enable a lower threshold on the hadronic activity. This increases the acceptance for SUEP signatures, which are predicted by hidden-valley models with a large 't Hooft coupling. The observed results are consistent with the standard model background prediction. The most stringent limits to date are set on the gluon fusion production of heavy scalar mediators resulting in SUEP-like signals.}

\hypersetup{%
pdfauthor={CMS Collaboration},%
pdftitle={Search for soft unclustered energy patterns in proton-proton collisions at sqrt(s) = 13 TeV using data scouting},%
pdfsubject={CMS},%
pdfkeywords={CMS, dark QCD, dark matter, hidden valley, exotica, SUEP, scouting}}

\maketitle

Hidden-valley models with a strongly coupled dark sector (DS)~\cite{Strassler:2006im} are a credible possibility for physics beyond the standard model (BSM) at the CERN LHC. These models can provide dark-matter candidates and realizations of neutral naturalness~\cite{Batell:2022pzc}---a solution to the hierarchy problem~\cite{Craig:2022eqo} involving new particles not charged under standard-model (SM) gauge groups. They can also lead to a wide range of unconventional signatures~\cite{Cohen:2015toa,Schwaller:2015gea,Knapen:2016hky,CMS:2024zqs}. The observable phenomena from collider production are determined by the coupling between the hidden and SM sectors and by the dynamics of the DS~\cite{Strassler:2006im}. Specifically, soft unclustered energy patterns (SUEPs), which consist of high multiplicities of diffusely distributed, low-momentum charged particles, are difficult to distinguish from SM backgrounds at the trigger level in current LHC experiments. This Letter presents a search for SUEPs from the decay of a scalar mediator produced by gluon-gluon fusion (ggF), using the data-scouting strategy~\cite{CMS:2024zhe} of the CMS experiment. With limited bandwidth, data scouting lowers trigger thresholds relative to standard data taking by recording only partial information for each event, namely the trigger-level reconstruction output, thereby increasing acceptance for SUEP signals.

At the LHC, DS particles can be produced via a mediator that couples to both the SM particles and the DS. The DS extends the SM with a new non-Abelian gauge group including dark quarks and dark gluons, analogous to the SM quarks and gluons, with behavior similar to the SM quantum chromodynamics (QCD). If the dark-quark masses are below the dark-confinement scale, $\lambdaD$, and $\lambdaD \ll \sqrt{s}$, the dark quarks will shower and hadronize to dark hadrons, which may subsequently decay to SM particles. Typical models assume collinear emissions leading to collimated jets similar to those in the SM QCD. However, if the DS possesses a large 't Hooft coupling ($\lambda \equiv \alphaD \Ncdark \gg 1$, where \alphaD is the dark QCD coupling constant and \Ncdark is the number of dark colors), then large-angle emissions are unsuppressed for energies above the confinement scale~\cite{Knapen:2016hky}. In this case, DS particles will be emitted isotropically in the center-of-mass frame of the shower without large hierarchies in energy. This process is analogous to hadron production in QCD at high temperatures~\cite{Shuryak:2004cy,Schafer:2005ff,Kovtun:2004de,CasalderreySolana:2011us}. If there is a large mass difference between the mediator and the dark hadrons, radiation is efficient over a large energy range during the showering, resulting in a large multiplicity of low-momentum particles, the lightest states in the DS spectrum~\cite{Strassler:2008bv}. The distribution of the dark-hadron momenta can be described by a relativistic Boltzmann distribution~\cite{Barron:2021btf}, characterized only by the dark-hadron mass and a temperature parameter related to the confinement scale. Similar signatures of large multiplicity and isotropic events are predicted in scenarios with microscopic black holes in models with large extra spatial dimensions~\cite{Park:2012fe}.

\begin{figure}[!hbtp]
	\centering
	\includegraphics[width=\figWidth]{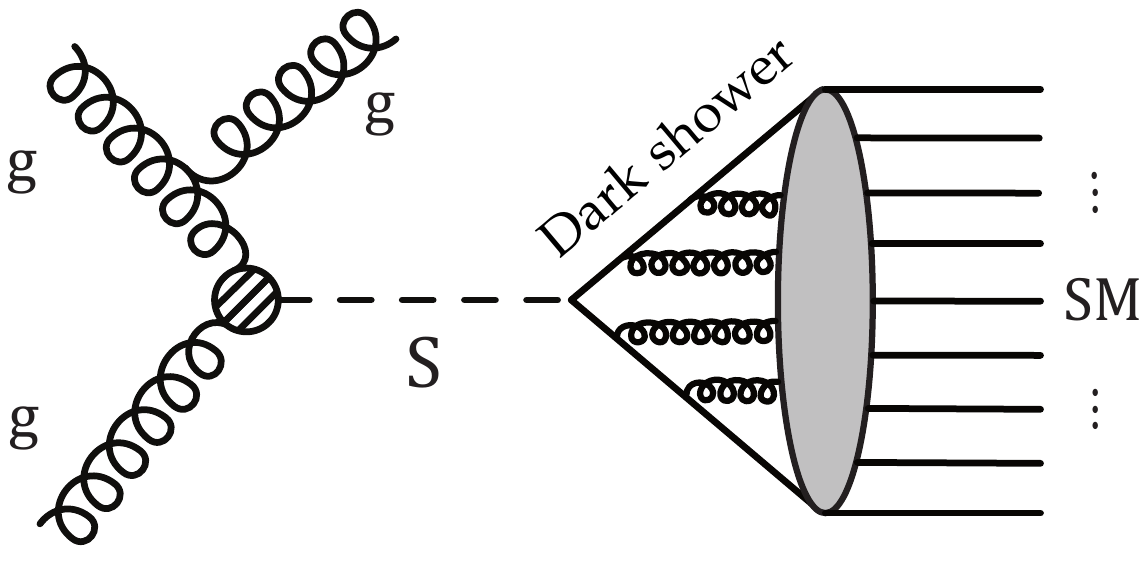}
	\caption{A schematic diagram of the SUEP signal production with initial-state radiation.}
	\label{fig:diagram}
\end{figure}

This search targets a broad range of experimentally accessible variations of the characteristic SUEP signature. These variations are encompassed by simplified models with a few parameters that control the kinematic behavior of the SUEP cluster. In the simplified models, the DS is accessed by a heavy scalar mediator \pS produced via ggF through an effective coupling, $\pS G_{\mu\nu} \PXXG^{\mu\nu}$ or $\pS G_{\mu\nu} \widetilde{\PXXG}^{\mu\nu}$, and assumed to decay to a SUEP with 100\% branching fraction. The nominal signal cross sections correspond to BSM Higgs boson production via a point-like effective interaction~\cite{LHCHiggsCrossSectionWorkingGroup:2016ypw} compatible with experimental limits. The calculation includes next-to-next-to-next-to-leading order QCD corrections, but no additional electroweak or off-shell production effects, and uses the PDF4LHC15\_nnlo\_100 set of parton distribution functions~\cite{Carrazza:2015aoa}. The showering and hadronization in the DS are modeled as the emission of low-mass dark \pphi mesons, with momenta sampled from a relativistic Boltzmann distribution that depends on the dark-meson mass \mPhi and temperature \TD at the confinement scale \lambdaD. High-multiplicity low-momentum emissions are obtained by requiring \mPhi and \TD to be at the same energy scale of a few {\GeVns}. The dark mesons decay into a pair of dark photons \PApr via a $\pphi \PApr^{\mu \nu}\PApr_{\mu \nu}$ operator. Through kinetic mixing with the SM hypercharge gauge boson, the dark photon promptly decays into a pair of SM electrons, muons, or charged pions: $\PApr \to \EE$, \MM, or $\pipi$. The result is a fully visible, prompt SUEP event, as shown in Fig.~\ref{fig:diagram}.

Simulated SUEP events are generated with a custom package~\cite{Knapen:2016hky} interfaced with the \PYTHIA 8.240~\cite{Sjostrand:2014zea} generator for mediator masses \mS ranging from 125\GeV to 2\TeV, dark-meson masses \mPhi from 1 to 8\GeV, and temperatures \TD from $\mPhi/4$ to $4\mPhi$. Three dark-photon scenarios, with different masses \mApr and corresponding \EE, \MM, and \pipi decay modes, are considered: $\mApr = 0.5\GeV$ and lepton-dominated decays with branching fractions $\BF = 40$, 40, and 20\%, respectively; $\mApr = 0.7\GeV$ and hadron-dominated decays with $\BF = 15$, 15, and 70\%, respectively; and $\mApr = 1\GeV$ with only fully hadronic decays $\PApr \to \pipi$. The mediator mass, dark-meson mass, temperature, and dark-photon decay mode collectively determine the multiplicity, energy spectrum, and flavor composition of the constituents of the SUEP cluster. The \PYTHIA generator with the CP5 underlying-event tune~\cite{CMS:2019csb} is also used to simulate multiple proton-proton (\pp) interactions (pileup). The \GEANTfour software~\cite{Agostinelli:2002hh} is used to simulate particle interactions with the CMS detector.

The CMS apparatus~\cite{CMS:2008xjf,CMS:2023gfb} is a multipurpose, nearly hermetic detector, designed to reconstruct and identify electrons, muons, photons, and hadrons~\cite{CMS:2020uim,CMS:2018rym,CMS:2014pgm}. A global ``particle-flow'' (PF) algorithm~\cite{CMS:2017yfk} aims to reconstruct all individual particles in an event, combining information provided by the all-silicon inner tracker and by the crystal electromagnetic and brass-scintillator hadron calorimeters, operating inside a 3.8\unit{T} superconducting solenoid, with data from the gas-ionization muon detectors embedded in the flux-return yoke outside the solenoid. The reconstructed particles are used to build jets and missing transverse momentum~\cite{CMS:2016lmd,CMS:2019ctu}.
Events of interest are selected using a two-tiered trigger system. The first level (L1), composed of custom hardware processors, uses information from the calorimeters and muon detectors to select events at a rate of around 100\unit{kHz} within a fixed latency of about 4\mus~\cite{CMS:2020cmk}. The second level, known as the high-level trigger (HLT), consists of a farm of processors running a version of the full event reconstruction software optimized for fast processing, and reduces the event rate to around 1\unit{kHz} before data processing and storage for the standard physics program~\cite{CMS:2024aqx}, as well as around 5\unit{kHz} each for the scouting and parking streams~\cite{CMS:2016ngn}.

The data analyzed in this Letter correspond to an integrated luminosity of \scoutinglumi from \pp collisions at $\sqrt{s} = 13\TeV$ collected with the CMS detector during 2016--2018. The data were recorded by requiring high hadronic activity at L1 and HLT, measured by \HTHLT, the scalar sum of the transverse momentum (\pt) of calorimeter jets with $\pt>40\GeV$ and pseudorapidity $\abs{\eta} < 2.5$. The high-\HTHLT trigger targets SUEP signals with significant initial-state radiation (ISR). The first CMS search for SUEPs from ggF production~\cite{CMS:2024nca} used the data collected by high-\HT triggers with thresholds around 1\TeV. These data were analyzed with the full ``offline'' reconstruction. The CMS experiment has also searched for SUEPs from a Higgs boson decay, produced in association with a vector boson~\cite{CMS:2026jia}. The ATLAS experiment has searched for SUEPs containing muons in the final state~\cite{ATLAS:2026fso}. The search described here uses a scouting trigger with a lower \HTHLT threshold than in the previous ggF search. This is accomplished by storing only the reconstructed objects---PF candidates, PF jets, and primary vertices---from the HLT fast processing, discarding the raw event data and performing no further offline reconstruction.
Therefore, while less information is available for data analysis, the event size is reduced by a factor of ${\approx}100$, which enables decreasing the threshold on \HTHLT from 1\TeV to 410\GeV.
This increases the trigger efficiency by up to one order of magnitude for SUEP models with a wide range of parameters, especially for \TD and \mPhi $\lesssim 2\GeV$, which result in a higher fraction of unclustered energy. 

Primary vertices (PVs) are reconstructed by clustering tracks based on their distance of closest approach to the \pp interaction point along the $z$ axis. The vertex with the largest $\Sigma \pt^2$ of associated tracks is identified as the PV. Charged PF candidates are required to have $\pt >0.75\GeV$ and $\abs{\eta} < 2.4$, and only those associated with the PV are retained in order to distinguish high-multiplicity, isotropic signal events from pileup collisions. The per-particle reconstruction efficiency for the signal ranges from 45 to 95\%, depending on the charged-particle \pt. This is lower than the ${\approx}90\%$ offline efficiency because of the simplified tracking algorithm used to satisfy the HLT computational constraints~\cite{CMS:2024zhe}. Jets are clustered from PF candidates using the anti-\kt (AK) algorithm with a distance parameter of $R = 0.4$ (AK4)~\cite{Cacciari:2008gp,Cacciari:2011ma} and are required to have $\pt > 30\GeV$ and $\abs{\eta} < 2.6$.

\ifthenelse{\boolean{cms@external}}{%
\begin{figure*}[!hbtp]
	\centering\includegraphics[width=0.99\textwidth]{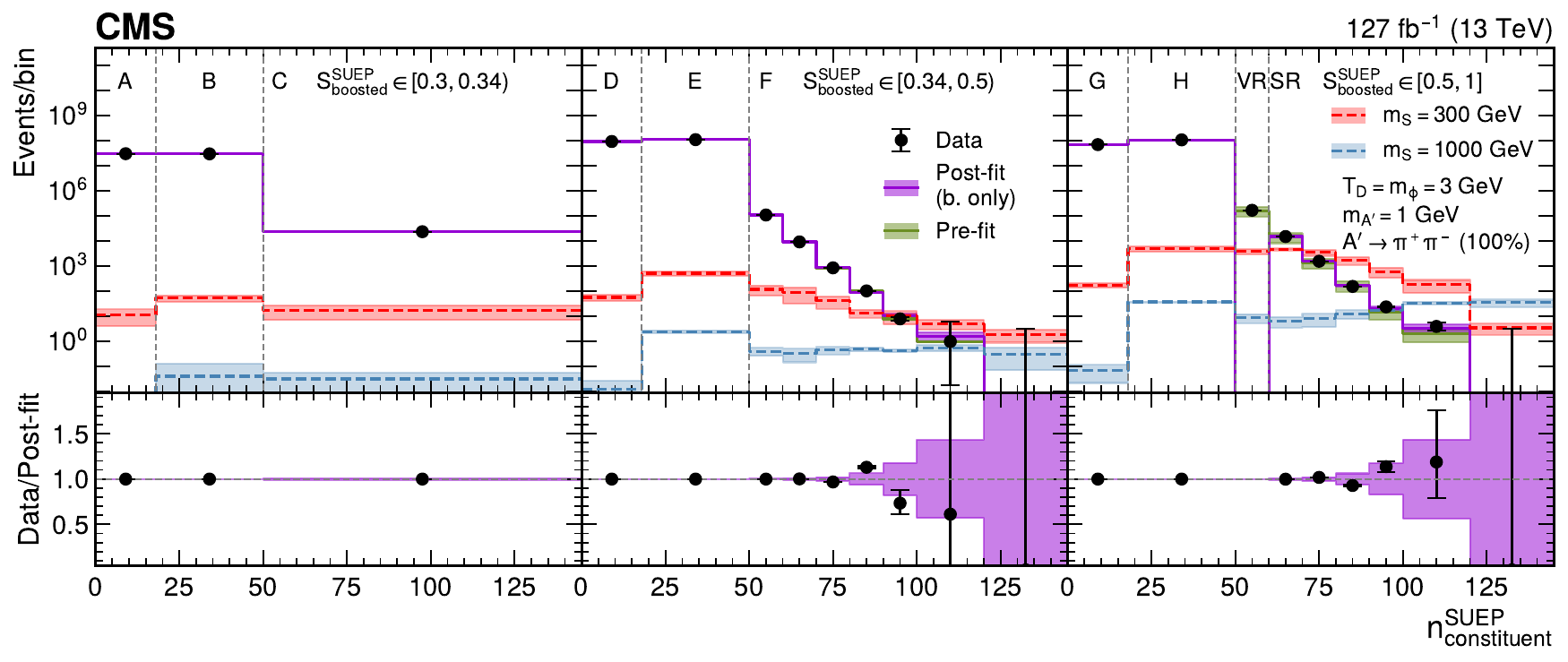}
	\caption{The observed \nSUEPConstituents distributions for various ranges of \boostedSphericity (left: $0.3 < \boostedSphericity < 0.34$, middle: $0.34 < \boostedSphericity < 0.5$, right: $\boostedSphericity > 0.5$), compared to the background prediction in the SR. The pre-fit expected background from Eq.~\eqref{eq:extended-ABCD}, the background-only fit result, and two signal models with $\mS = 300$ and 1000\GeV are shown, as well as their uncertainties. Both signal models have $\TD = \mPhi = 3\GeV$ and $\mApr = 1\GeV$ (fully hadronic decays).
	The vertical dashed lines mark the boundaries that separate the CRs, the VR, and the SR. The lower panels show the ratio of data to the post-fit background prediction.}
	\label{fig:regions-postfit}
\end{figure*}
}{}

Events that pass the trigger, basic data quality criteria, and the $\HT>560\GeV$ requirement, based on the \HT calculated as a \pt sum of the AK4 PF jets, which ensures the trigger is fully efficient, are preselected. The SUEP final state particles are boosted to form a jet, with a radius that scales as $\mS/\pt^{\pS}$, recoiling against the ISR system. Charged PF candidates are clustered into wide AK jets with a distance parameter of $R=1.5$ (AK15) and $\pt > 50\GeV$. The $R$ value is chosen to contain as much of the SUEP activity as possible, while preserving separation from the ISR system. The SUEP-ISR system is defined by the two AK15 jets with the largest \pt. The jet with the larger number of charged PF candidates, hereafter denoted constituents, is identified as the SUEP candidate, while the other jet is treated as the ISR candidate. To recover the isotropic event shape, the constituents of the reconstructed SUEP-candidate AK15 jet are boosted into the jet rest frame and used to compute the generalized sphericity tensor $\SAB = \Sigma_{i}\abs{\momi}^{-1}\momi^{\alpha}\momi^{\beta} / \Sigma_{i}\abs{\momi}$, where $\alpha,\beta$ are the Euclidean coordinates and \momi is the momentum of the constituent $i$~\cite{Cesarotti:2020ngq}. The boosted sphericity $\boostedSphericity = \frac{3}{2} (\lambda_{2} + \lambda_{3})$, where $\lambda_{2,3}$ are the two smallest \SAB eigenvalues, ranges from 0, indicating a two-prong structure, to 1, indicating spherical symmetry.

\ifthenelse{\boolean{cms@external}}{}{%
\begin{figure*}[!hbtp]
	\centering\includegraphics[width=0.99\textwidth]{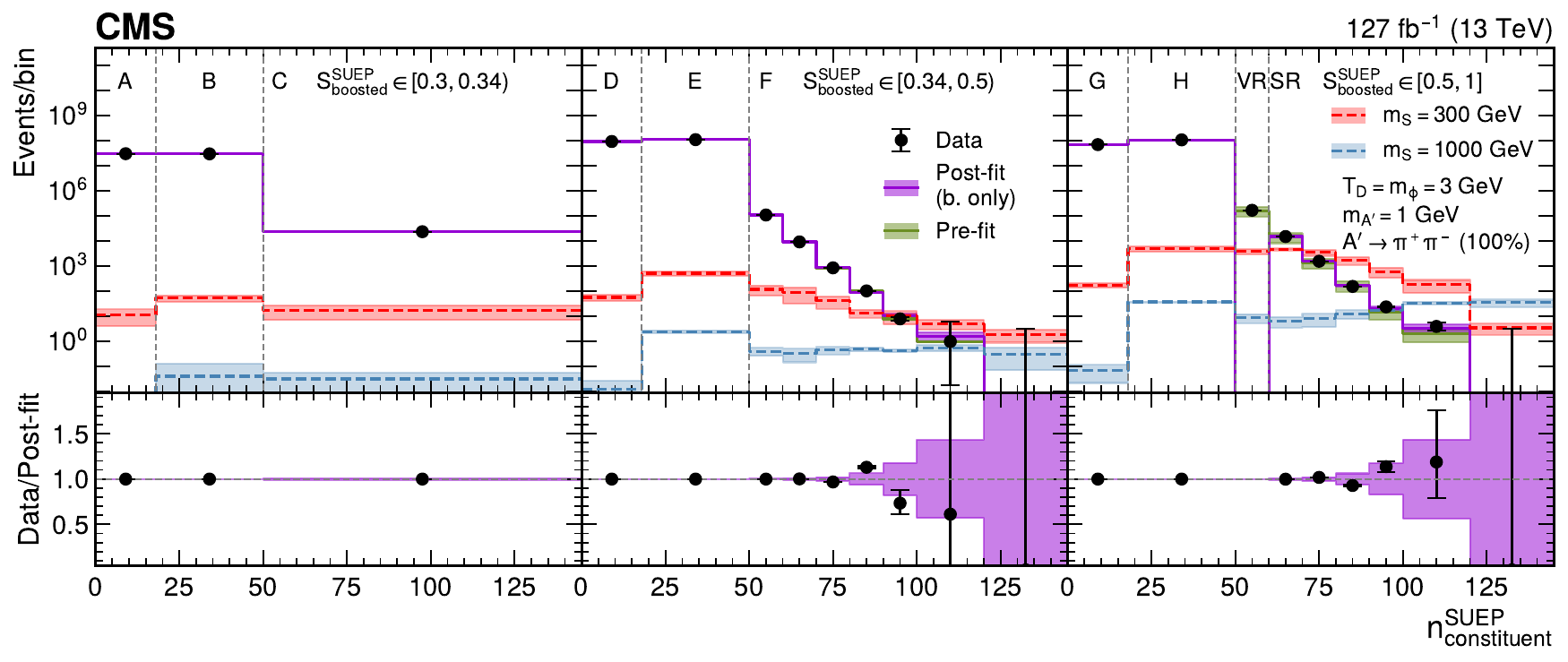}
	\caption{The observed \nSUEPConstituents distributions for various ranges of \boostedSphericity (left: $0.3 < \boostedSphericity < 0.34$, middle: $0.34 < \boostedSphericity < 0.5$, right: $\boostedSphericity > 0.5$), compared to the background prediction in the SR. The pre-fit expected background from Eq.~\eqref{eq:extended-ABCD}, the background-only fit result, and two signal models with $\mS = 300$ and 1000\GeV are shown, as well as their uncertainties. Both signal models have $\TD = \mPhi = 3\GeV$ and $\mApr = 1\GeV$ (fully hadronic decays).
	The vertical dashed lines mark the boundaries that separate the CRs, the VR, and the SR. The lower panels show the ratio of data to the post-fit background prediction.}
	\label{fig:regions-postfit}
\end{figure*}
}

Events that pass the preselection are further required to have the number of constituents in the SUEP candidate \nSUEPConstituents greater than 50 and $\boostedSphericity > 0.5$. These criteria significantly suppress most SM background processes including top quark production, leaving primarily QCD multijet events with high track multiplicity. The remaining background is estimated using an extended ABCD method~\cite{Choi:2019mip}, which relies on two discriminating variables without large statistical dependence: \nSUEPConstituents and \boostedSphericity. These variables define a phase space partitioned into one signal region (SR) and multiple control regions (CRs) based on selection thresholds. Unlike the standard ABCD method, which divides the phase space using one threshold per variable to create four regions (three CRs and one SR), the extended ABCD approach introduces two thresholds for each variable, resulting in a $3\times3$ grid of nine exclusive regions. This extension improves the treatment of correlations between \nSUEPConstituents and \boostedSphericity in background events.
The definitions of the CRs, labeled A--H, and the SR are shown in Fig.~\ref{fig:regions-postfit}, along with the corresponding distributions of data and two benchmark signal models with $\mS = 300$ and 1000\GeV, $\TD = \mPhi = 3\GeV$, and $\mApr = 1\GeV$ (fully hadronic decays). The SR selection efficiency can fall below 5\% for signals with low \mS or high \TD and \mPhi, while exceeding 90\% for signals with large \mS. To enable a differential background estimation, both the SR and region F are further subdivided into bins of $\nSUEPConstituents$: $[50,60)$, $[60, 70)$, $[70, 80)$, $[80, 90)$, $[90, 100)$, $[100, 120)$, and $> 120\GeV$.

Assuming that the \nSUEPConstituents distribution in the SR has a similar shape to that in region F, which is adjacent to the SR in \boostedSphericity, the background yield in a given \nSUEPConstituents bin of the SR, \NSRbin, is given by:
\begin{equation}\label{eq:extended-ABCD}
	\NSRbin = \corrF \NFbin \NSR,\; \NSR = \frac{\NF \NH^2 \ND^2 \NB^2}{\NGg \NC \NAa \NE^4},
\end{equation}
with $N_X$ being the event yield in region $X$ and \NFbin the yield in the corresponding \nSUEPConstituents bin in region F. The factor \corrF corrects for higher-order correlations, accounting for deviation from the matrix method assumption of $(\NSRbin/\NSR)/(\NFbin/\NF) = 1$. The \corrF values are extracted from a linear fit to the ratio of the normalized \nSUEPConstituents distributions in CRs F and C to suppress the statistical fluctuations, as detailed in \EndMatterA.

The uncertainty in the per-bin distribution $\NSRbin/\NSR$ arises from higher-order correlations between \boostedSphericity and \nSUEPConstituents not captured by \corrF. This uncertainty is estimated by propagating the statistical uncertainties in the F and C bins in the \corrF fit, separately for each data-taking year and correlated across the SR bins, assuming smooth \nSUEPConstituents distributions. A 100\% uncertainty is assigned to the last bin $\nSUEPConstituents > 120$. A separate normalization uncertainty for the SR bins in each data-taking year, also arising from higher-order correlations, is assessed by applying the extended ABCD method to the discriminating variables from the ISR candidate jet, with the magnitude of 50\% taken from the resulting disagreement between the prediction and the observation. When fitting the statistical model to the observed data, we allow for the normalization factor to float in the fit within the above uncertainty. This uncertainty is found to be constrained after the fit by the observed yields in low-\nSUEPConstituents bins, which dominate the total SR yield and contain much more background than potential signal contributions. Therefore, a high sensitivity to SUEP signals is still achieved with this yield uncertainty.
 
The background estimation method is validated by applying the approach to simulated QCD multijet events and comparing the predictions to the values directly taken from the simulation in the SR. The method is further validated in data using the first SR bin $[50,60)$ as the validation region (VR), as it has negligible signal contributions and is adjacent to the other SR bins, to test the agreement between the predicted and observed yields and distributions. Details of the validation are given in \EndMatterB.

The signal simulation modeling is subject to systematic uncertainties in the trigger efficiency, the AK4 jet energy scale propagated to \HT, the pileup distribution, the trigger prefiring probability in 2016--2017~\cite{CMS:2020cmk}, the integrated luminosity measurement~\cite{CMS:2021xjt,CMS:2018elu,CMS:2019jhq}, and the parton shower modeling. For the $\mS = 125\GeV$ signal sample with the same \mS and production mechanism as the Higgs boson, higher-order corrections for the \pS kinematic behavior~\cite{Becker:2020rjp} are applied as a \pt-dependent reweighting with a corresponding uncertainty. The dominant uncertainty is in the tracking efficiency, up to 50\% for the highest \nSUEPConstituents SR bins, while other uncertainties are small. The tracking efficiency uncertainty is estimated by randomly removing tracks based on the difference in the track reconstruction efficiency for \PDst mesons in data and simulation~\cite{trackeff,cms2712742muon}. These probabilities, measured for the offline tracks, are scaled up by a factor of ${\approx}3$ to account for the differences between the scouting and offline tracking efficiency uncertainties, derived from the variations of the efficiency across data-taking periods.

\begin{figure*}[!hbtp]
	\centering
	\includegraphics[width=\figWidthThree]{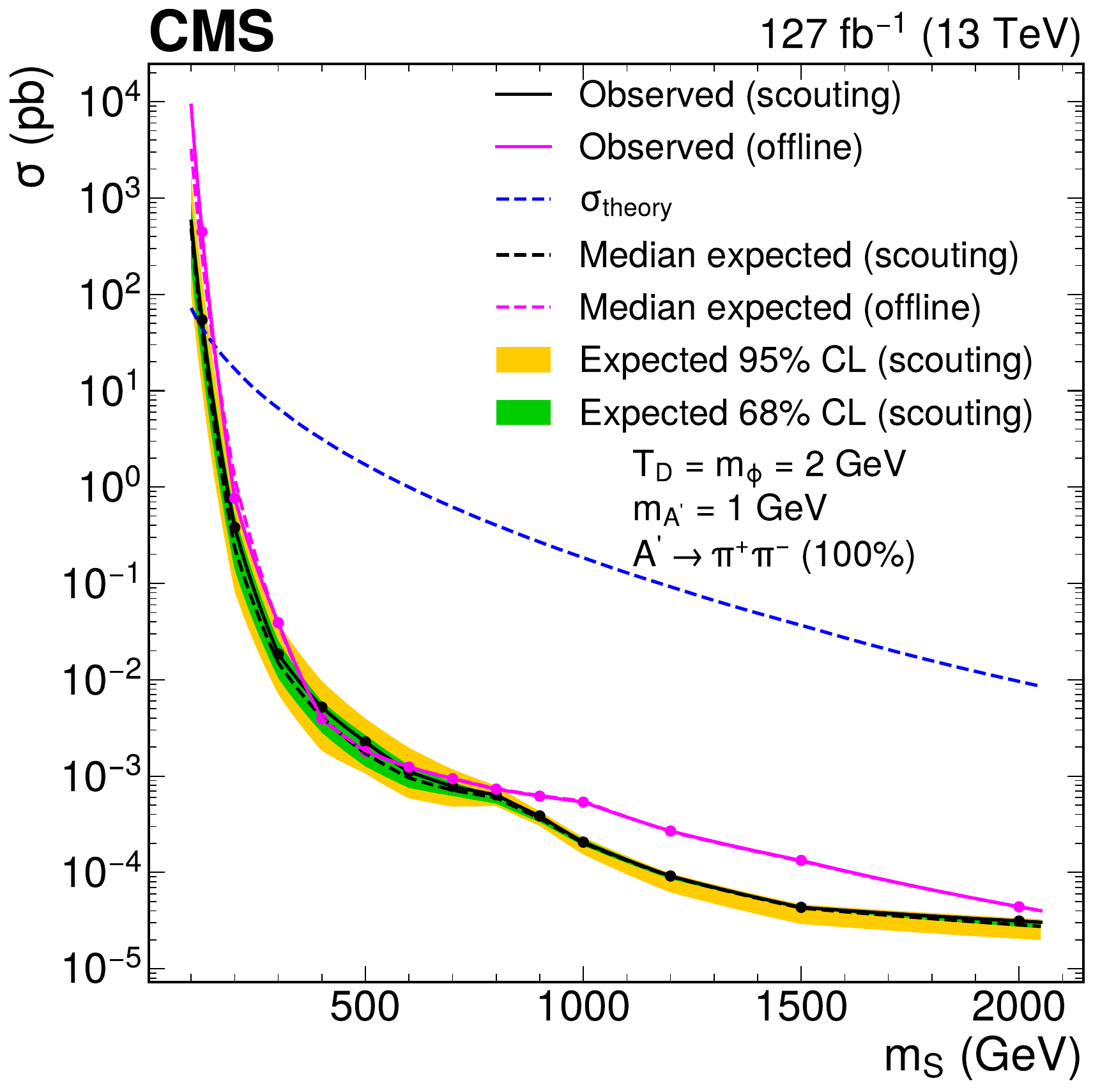}
	\includegraphics[width=\figWidthThree]{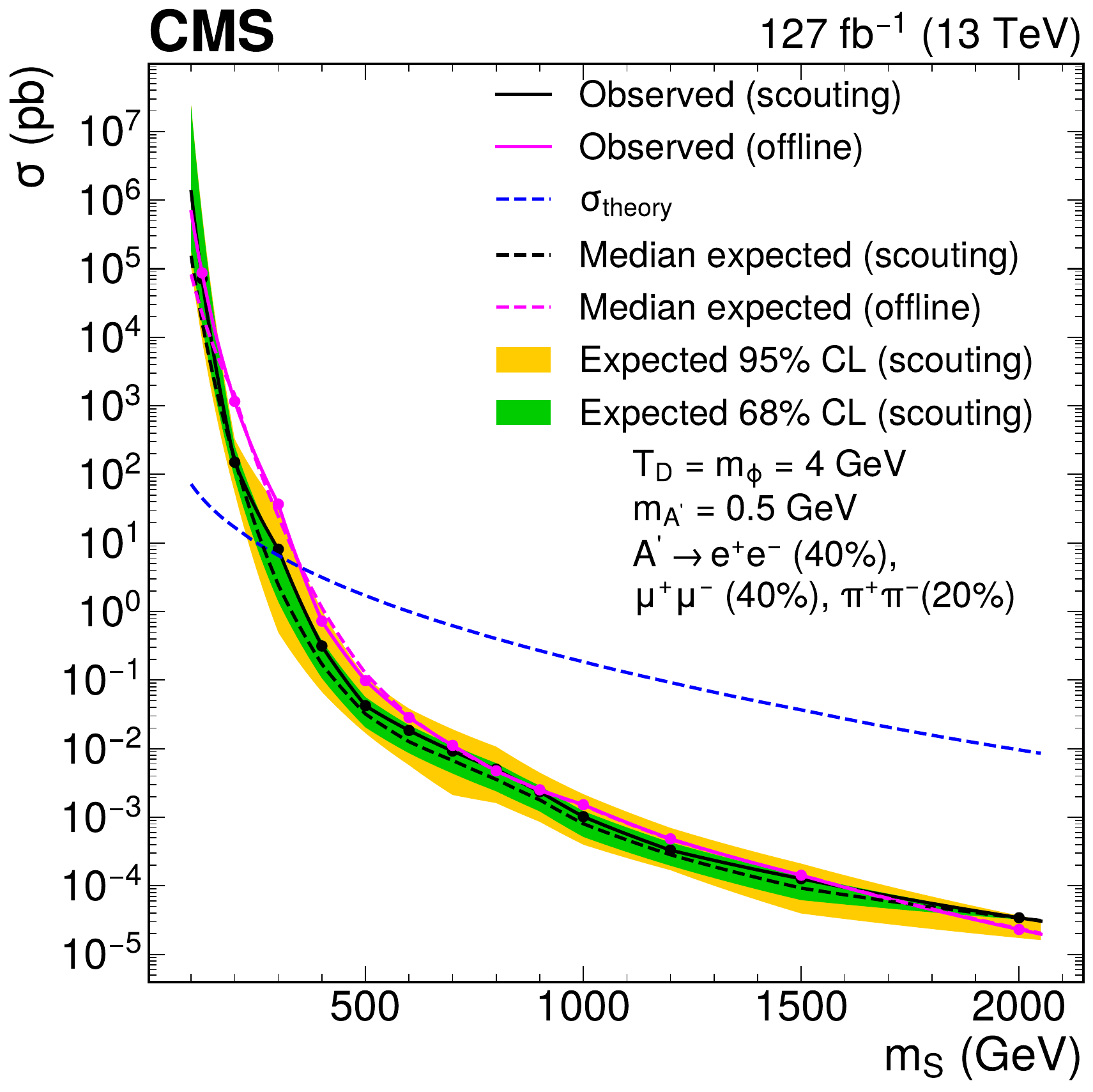}
	\includegraphics[width=\figWidthThree]{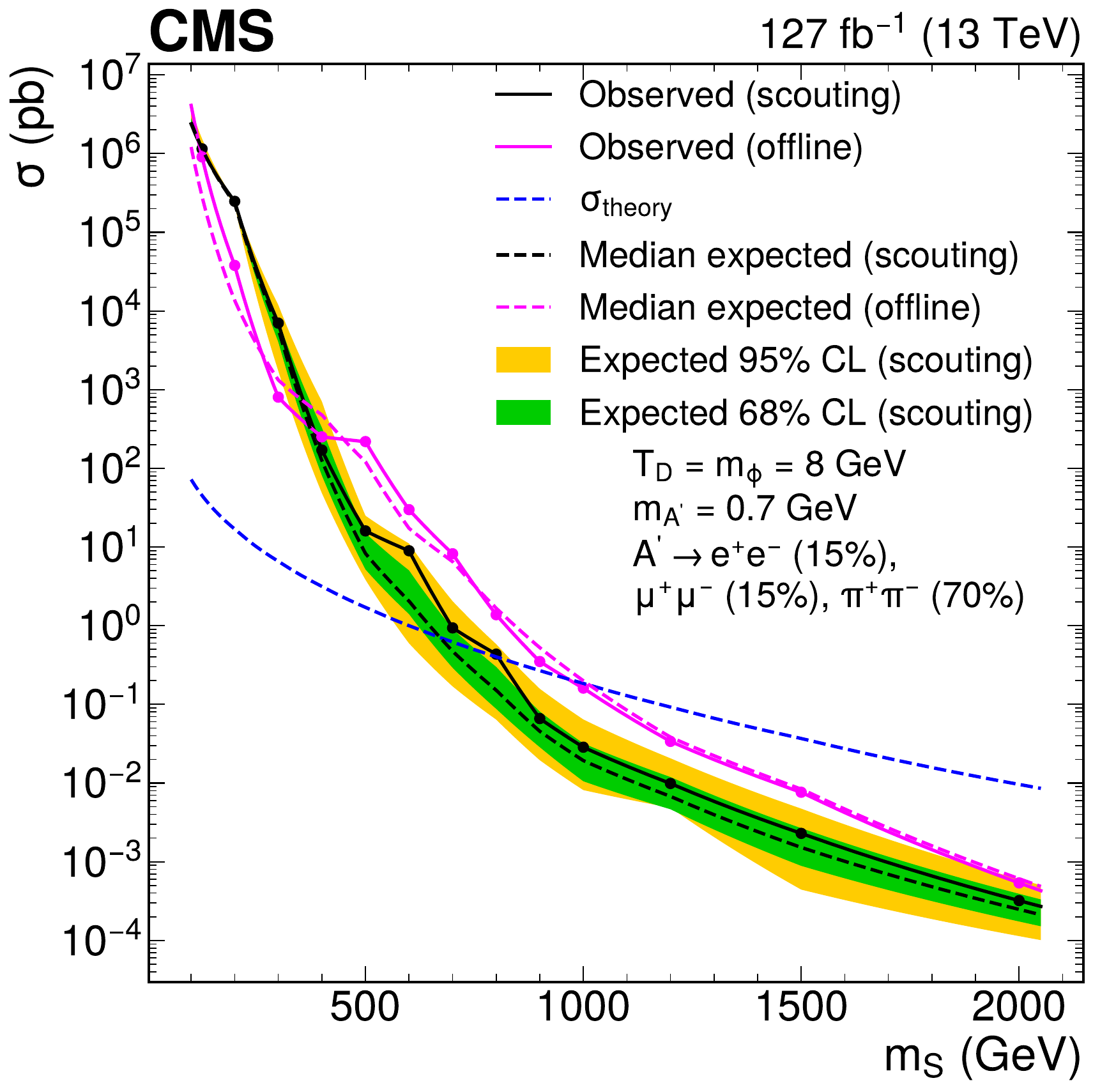}
	\caption{Expected and observed 95\% CL upper limits on the simplified scalar-mediator SUEP cross sections as functions of the mediator mass \mS, with a comparison to the corresponding limits from the previous CMS offline search~\cite{CMS:2024nca}. Signal models with different values of the temperature, dark hadron mass, and dark photon mass are shown. \cmsUpperLeft: $\TD = \mPhi = 2\GeV$, $\mApr = 1\GeV$ (fully hadronic decays). \cmsUpperRight: $\TD = \mPhi = 4\GeV$, $\mApr = 0.5\GeV$ (lepton-dominated decays). \cmsLower: $\TD = \mPhi = 8\GeV$, $\mApr = 0.7\GeV$ (hadron-dominated decays).}
	\label{fig:exclusions}
\end{figure*}

\begin{figure*}[!hbtp]
	\centering
	\includegraphics[width=0.99\textwidth]{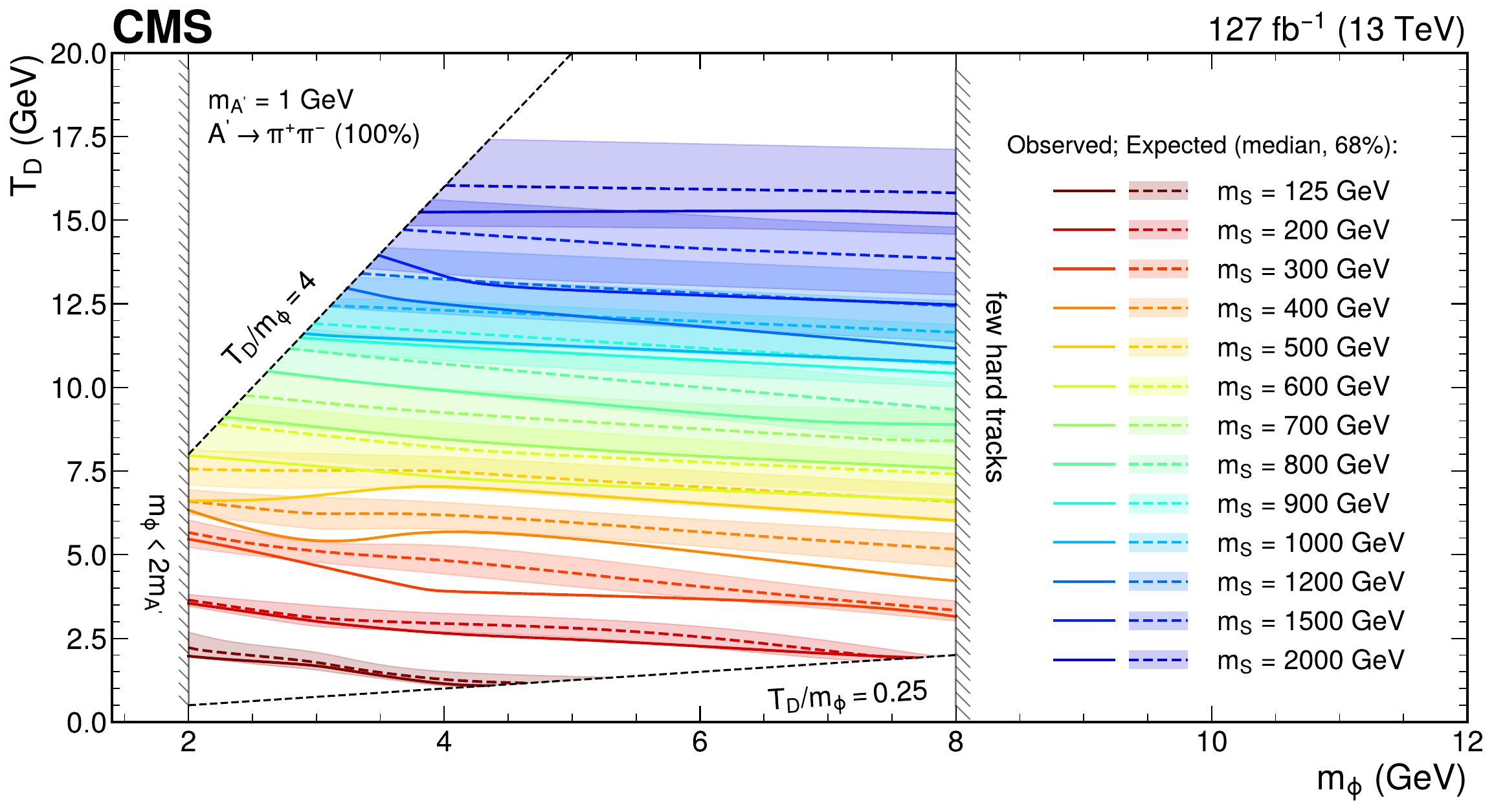}
	\caption{Expected and observed 95\% CL exclusion limits in the $(\TD,\mPhi)$ plane, assuming ggF production of scalar mediators \pS with the nominal cross sections and $\mApr = 1\GeV$ (fully hadronic decays). The region below each curve is excluded for that signal model.}
	\label{fig:exclusionsscan}
\end{figure*}

A binned maximum likelihood fit~\cite{CMS:2024onh} is performed to the \nSUEPConstituents distribution, including the CRs and SR, but excluding the VR, treating the three data-taking years separately.
The expected central value, the sum of background and signal, in each bin or region follows a Poisson distribution. The expected background yields in the SR bins are constrained by Eq.~\eqref{eq:extended-ABCD}, while the CR yields are unconstrained. The expected signal distributions are taken from simulation, and the normalization is determined by the fit. The observations are consistent with the SM background predictions. The background-only fit results are shown in Fig.~\ref{fig:regions-postfit}. The 95\% confidence level (CL) upper limits on the signal production cross section are computed using the modified frequentist \CLs~\cite{Junk:1999kv,Read:2002hq} approach with pseudo-experiments (Fig.~\ref{fig:exclusions}) and an asymptotic approximation~\cite{Cowan:2010js} (Fig.~\ref{fig:exclusionsscan}), with similar performance. Limits as functions of the mediator mass for several signal models are shown in Fig.~\ref{fig:exclusions} and compared to the limits set by an earlier search based on offline reconstruction~\cite{CMS:2024nca}. The present search improves the earlier constraints by as much as one order of magnitude in most of the parameter space, except for a small region with low \nSUEPConstituents, where the lower scouting track reconstruction efficiency reduces the sensitivity.
Figure~\ref{fig:exclusionsscan} presents the exclusion limits as functions of \TD and \mPhi, for models with different \mS values for fully hadronic \PApr decays. The limits for signal models with leptonic \PApr decays are comparable, since they exhibit similar kinematic properties and the analysis does not explicitly exploit particle flavor information. Tabulated results are available in the HEPData record for this analysis~\cite{HEPData}.

In summary, a search has been reported for soft unclustered energy patterns (SUEPs) at the LHC using scouting proton-proton collision data at $\sqrt{s}=13\TeV$, corresponding to an integrated luminosity of \scoutinglumi. This data set is recorded by a high-level trigger with a lower threshold on the scalar sum (\HT) of jet transverse momenta (\pt) compared to the standard \HT triggers, offsetting the increase in rate by saving limited event content. The characteristic isotropic event shape of SUEP signals is recovered by analyzing the jet in the SUEP-candidate rest frame and selecting particles from only the SUEP candidate, which is chosen as the jet with the highest constituent multiplicity out of the two highest \pt large-radius jets. The standard model background from quantum chromodynamics multijet processes is estimated from data control regions. The search sets the most stringent limits on a large range of SUEP models with a temperature and dark-hadron mass around a few \GeV, and scalar mediator masses above 125\GeV.

\begin{acknowledgments}
	We congratulate our colleagues in the CERN accelerator departments for the excellent performance of the LHC and thank the technical and administrative staffs at CERN and at other CMS institutes for their contributions to the success of the CMS effort. In addition, we gratefully acknowledge the computing centers and personnel of the Worldwide LHC Computing Grid and other centers for delivering so effectively the computing infrastructure essential to our analyses. Finally, we acknowledge the enduring support for the construction and operation of the LHC, the CMS detector, and the supporting computing infrastructure provided by the following funding agencies: SC (Armenia), BMFWF and FWF (Austria); FNRS and FWO (Belgium); CNPq, CAPES, FAPERJ, FAPERGS, and FAPESP (Brazil); MES and BNSF (Bulgaria); CERN; CAS, MoST, and NSFC (China); MINCIENCIAS (Colombia); MSES and CSF (Croatia); RIF (Cyprus); SENESCYT (Ecuador); ERC PRG and PSG, TARISTU24-TK10 and MoER TK202 (Estonia); Academy of Finland, MEC, and HIP (Finland); CEA and CNRS/IN2P3 (France); SRNSF (Georgia); BMFTR, DFG, and HGF (Germany); GSRI (Greece); MATE and NKFIH (Hungary); DAE and DST (India); IPM (Iran); SFI (Ireland); INFN (Italy); MSIT and NRF (Republic of Korea); MES (Latvia); LMTLT (Lithuania); MOE and UM (Malaysia); BUAP, CINVESTAV, CONACYT, LNS, SEP, and UASLP-FAI (Mexico); MOS (Montenegro); MBIE (New Zealand); PAEC (Pakistan); MSHE, NSC, and NAWA (Poland); FCT (Portugal);  MESTD (Serbia); MICIU/AEI and PCTI (Spain); MOSTR (Sri Lanka); Swiss Funding Agencies (Switzerland); MST (Taipei); MHESI (Thailand); TUBITAK and TENMAK (T\"{u}rkiye); NASU (Ukraine); STFC (United Kingdom); DOE and NSF (USA).
\end{acknowledgments}\section*{Data availability} Release and preservation of data used by the CMS Collaboration as the basis for publications is guided by the  \href{https://doi.org/10.7483/OPENDATA.CMS.1BNU.8V1W}{CMS data preservation, re-use and open access policy}.

\bibliography{auto_generated}

\appendix
\ifthenelse{\boolean{cms@external}}{\section*{\appname}}{\section{Additional details}}

\subsection{Corrections for the background estimation in the signal region}\label{app:correction}

In the extended ABCD method, the \nSUEPConstituents distribution in CR F is used as a proxy for the background shape in the SR. However, because of higher-order correlations between \nSUEPConstituents and \boostedSphericity, the SR background shape deviates from that observed in region F. This deviation is estimated by comparing the normalized \nSUEPConstituents distributions in regions F and C, motivated by the fact that regions C, F, and the SR share the same \nSUEPConstituents subdivisions and have adjacent \boostedSphericity ranges. Assuming only weak correlations between \nSUEPConstituents and \boostedSphericity, the trend of the SR-to-F shape ratio $(\NSRbin/\NSR)/(\NFbin/\NF)$ can be approximated by the F-to-C shape ratio, $(\NFbin/\NF)/(\NC^\text{bin}/\NC)$.

The estimated F-to-C ratios are then used as correction factors \corrF applied to the predicted SR background distribution in Eq.~\eqref{eq:extended-ABCD}. To mitigate statistical fluctuations in high-\nSUEPConstituents bins, where data are sparse, a linear fit is used to smooth the ratio as a function of constituent multiplicity. To ensure a stable fit, the \nSUEPConstituents histograms in regions F and C are constructed with finer binning, especially at low multiplicities where event yields are higher. The linear fit is performed on the bin-by-bin F-to-C ratios, weighted by the inverse squared statistical uncertainties in the ratios. The fit is carried out using the following \nSUEPConstituents bin boundaries: 50, 52, 54, 56, 58, 60, 62, 64, 66, 68, 70, 75, 80, and 120. The F-to-C ratios are fitted as a linear function of \nSUEPConstituents at the corresponding bin centers. The resulting linear function is then evaluated at the centers of bins with lower edges at 50, 60, 70, 80, 90, 100, and 120\GeV, used in the SR estimation, to obtain the \corrF values. For the last bin $[120, \infty)$, the correction is taken from the fit value at $\nSUEPConstituents = 125$.

This procedure is repeated separately for each data-taking year (2016, 2017, and 2018), and the corresponding year-specific corrections are applied to the prediction of the SR background distribution.

\subsection{Validations of the background estimation method}\label{app:validation}

\begin{figure*}[!hbtp]
	\centering
	\includegraphics[width=0.99\textwidth]{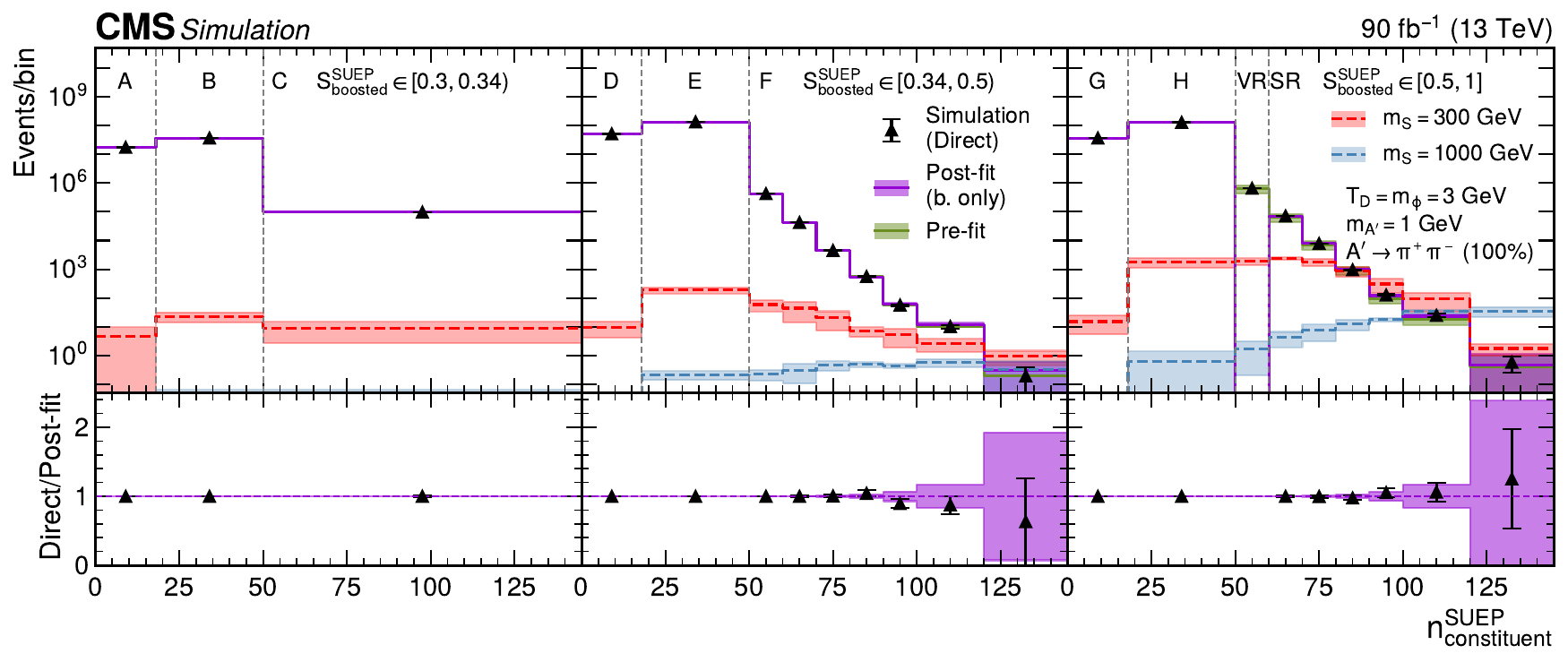}
	\caption{The \nSUEPConstituents distributions for various ranges of \boostedSphericity (left: $0.3 < \boostedSphericity < 0.34$, middle: $0.34 < \boostedSphericity < 0.5$, right: $\boostedSphericity \geq 0.5$) in QCD multijet simulation, comparing the extended ABCD prediction to the yield directly from the simulation in the SR. The pre-fit expected background from Eq.~\eqref{eq:extended-ABCD}, the background-only fit results, and two signal models with $\mS = 300$ and 1000\GeV are shown, as well as their uncertainties. Both signal models have $\TD = \mPhi = 3\GeV$ and $\mApr = 1\GeV$ (fully hadronic decays).
	The vertical dashed lines mark the boundaries that separate the CRs, the VR, and the SR. Simulated samples corresponding to the 2017 and 2018 data-taking conditions are used. The lower panels show the ratio of data to the post-fit background prediction.
	}\label{fig:closure_MC}
\end{figure*}

In Fig.~\ref{fig:closure_MC}, the extended ABCD background estimation is applied to the QCD multijet simulation for data-taking years 2017 and 2018, and the predicted background in the SR is compared to the event yields taken directly from the simulation in the SR. The results of the background-only fit are compatible with the simulation. The prediction from two signal models with $\mS = 300$ and 1000\GeV, $\TD = \mPhi = 3\GeV$, and $\mApr = 1\GeV$ (fully hadronic decays) are also shown. The signal distributions are distinct from the background.

\begin{figure*}[!hbtp]
	\centering
	\includegraphics[width=0.32\textwidth]{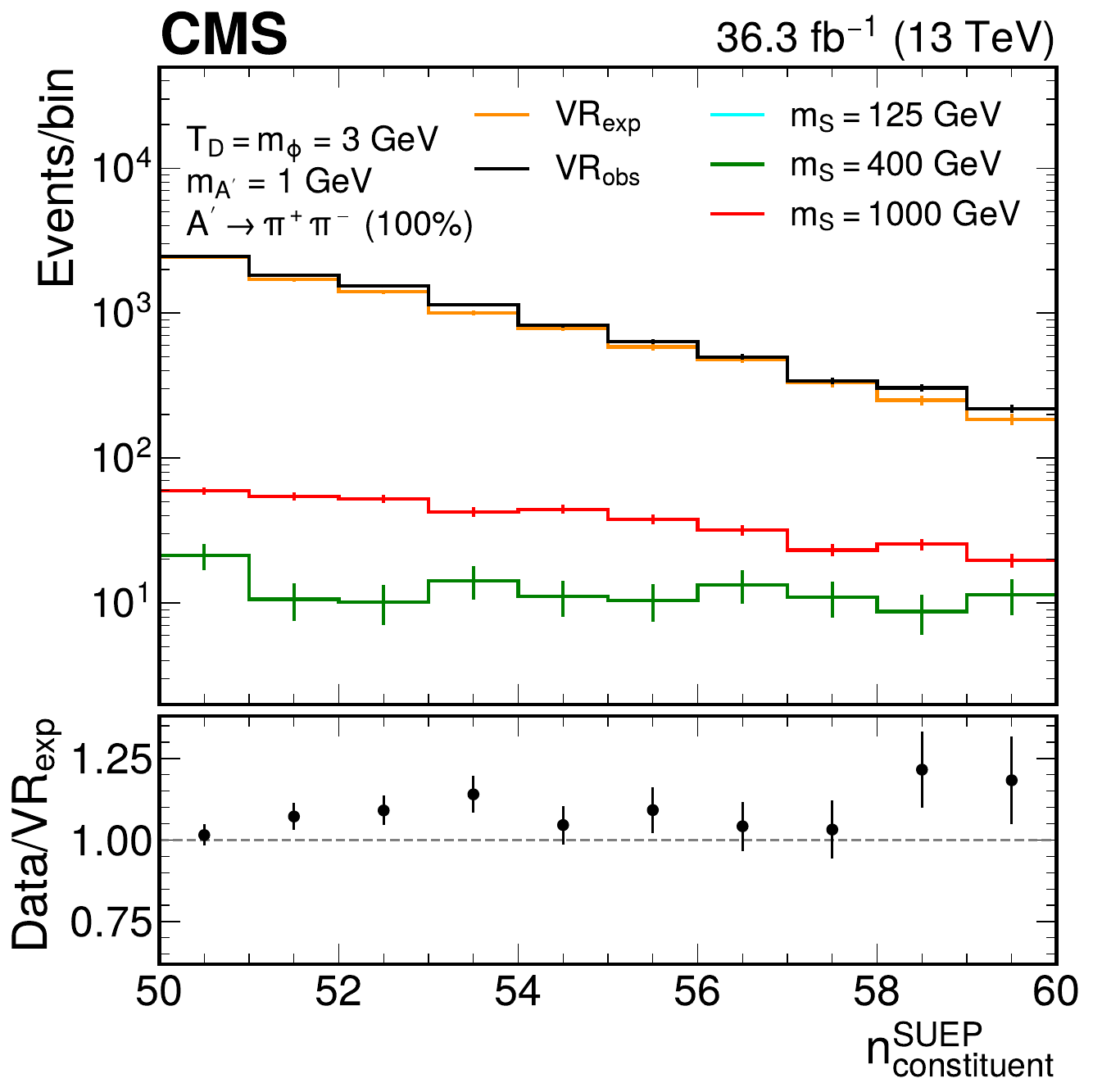}
	\includegraphics[width=0.32\textwidth]{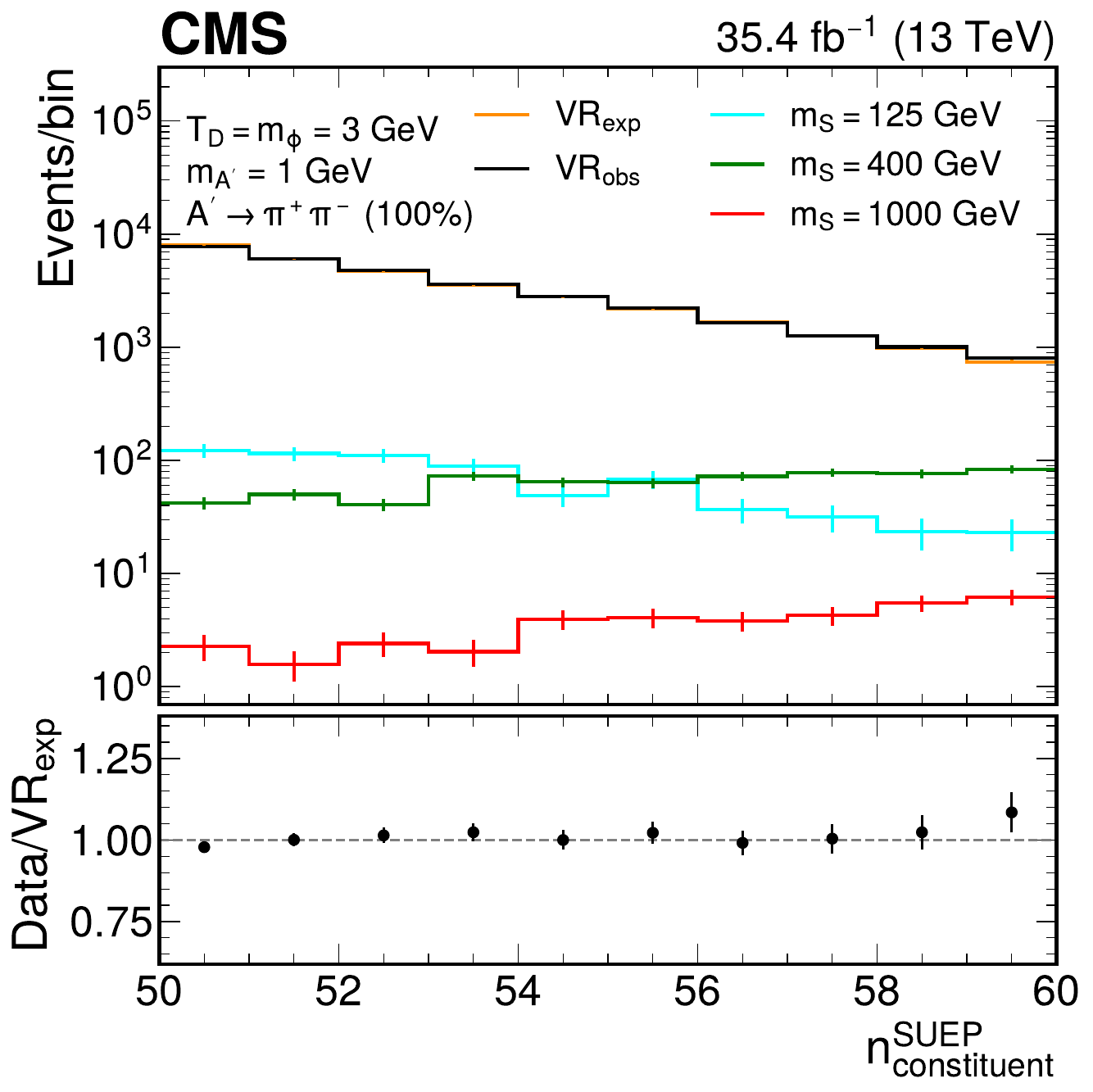}
	\includegraphics[width=0.32\textwidth]{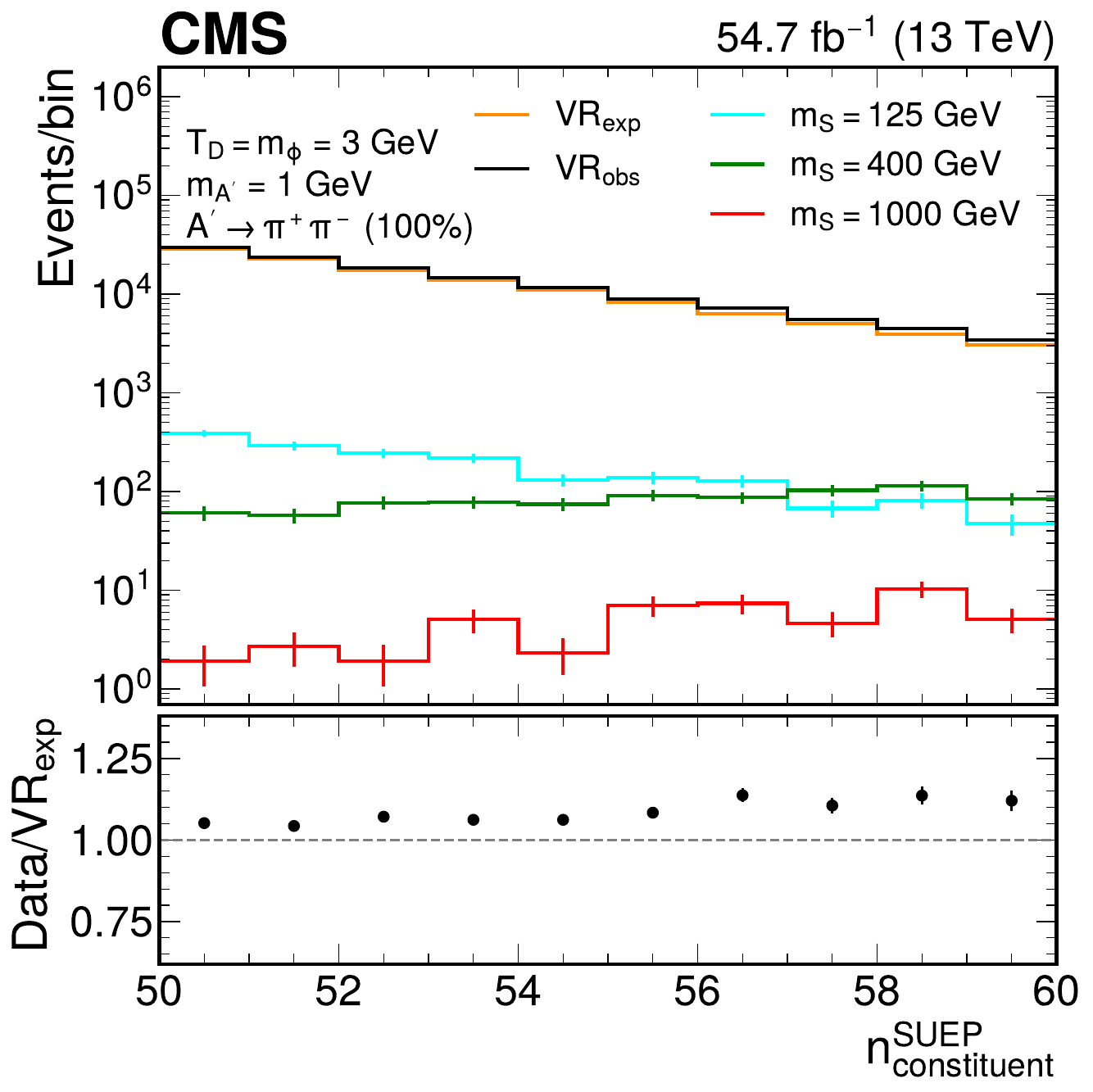}
	\caption{Validation of the extended ABCD method (left: 2016, middle: 2017, right: 2018) in the VR, where signal is negligible compared to background, as shown for signal models with $\mS = 125, 400\text{, and}~1000\GeV$, $\TD = \mPhi = 3\GeV$, and $\mApr = 1\GeV$ (fully hadronic decays). The lower panels show the ratio of data to the expected background in the VR.}
	\label{fig:closure_VR}
\end{figure*}

Figure~\ref{fig:closure_VR} shows the validation from applying the extended ABCD method to the VR: the first bin $\nSUEPConstituents \in [50,60)$ of the SR. The signal yield in this bin is small enough not to be visible above the background yield, as shown by the distributions of signal models with $\mS = 125, 400\text{, and}~1000\GeV$, $\TD = \mPhi = 3\GeV$, and $\mApr = 1\GeV$ (fully hadronic decays). The observed distributions are similar in shape to the prediction for all three data-taking years. The yield differences between the observation and the prediction are covered by the 50\% yield uncertainty, which is constrained by the background-dominated data in the fit.
\cleardoublepage \section{The CMS Collaboration \label{app:collab}}\begin{sloppypar}\hyphenpenalty=5000\widowpenalty=500\clubpenalty=5000\cmsinstitute{Yerevan Physics Institute, Yerevan, Armenia}
{\tolerance=6000
A.~Gevorgyan\cmsorcid{0000-0003-2751-9489}, A.~Hayrapetyan, V.~Makarenko\cmsorcid{0000-0002-8406-8605}, A.~Tumasyan\cmsAuthorMark{1}\cmsorcid{0009-0000-0684-6742}
\par}
\cmsinstitute{Marietta Blau Institute for Particle Physics, Vienna, Austria}
{\tolerance=6000
W.~Adam\cmsorcid{0000-0001-9099-4341}, L.~Benato\cmsorcid{0000-0001-5135-7489}, T.~Bergauer\cmsorcid{0000-0002-5786-0293}, M.~Dragicevic\cmsorcid{0000-0003-1967-6783}, S.~Gundacker\cmsorcid{0000-0003-2087-3266}, A.K.~Guven\cmsorcid{0009-0004-5670-5138}, P.S.~Hussain\cmsorcid{0000-0002-4825-5278}, M.~Jeitler\cmsAuthorMark{2}\cmsorcid{0000-0002-5141-9560}, N.~Krammer\cmsorcid{0000-0002-0548-0985}, A.~Li\cmsorcid{0000-0002-4547-116X}, D.~Liko\cmsorcid{0000-0002-3380-473X}, M.~Matthewman, J.~Schieck\cmsAuthorMark{2}\cmsorcid{0000-0002-1058-8093}, R.~Sch\"{o}fbeck\cmsAuthorMark{2}\cmsorcid{0000-0002-2332-8784}, M.~Shooshtari\cmsorcid{0009-0004-8882-4887}, M.~Sonawane\cmsorcid{0000-0003-0510-7010}, N.~Van~Den~Bossche\cmsorcid{0000-0003-2973-4991}, W.~Waltenberger\cmsorcid{0000-0002-6215-7228}, C.E.~Wulz\cmsAuthorMark{2}\cmsorcid{0000-0001-9226-5812}
% ...
\par}
$^{1}$Also at Yerevan State University, Yerevan, Armenia\\
$^{2}$Also at Technische Universit\"{a}t Wien, Vienna, Austria\\
% ...
\end{sloppypar}
\end{document}